\documentstyle[11pt,epsfig,epsf]{article}
\newcommand{\qv}{|\vec q|}
\newcommand{\be}{\begin{equation}}
\newcommand{\ee}{\end{equation}}
\newcommand{\bea}{\begin{eqnarray}}

\newcommand{\eea}{\end{eqnarray}}
\newcommand{\nn}{\nonumber}
\newcommand{\dd}{\displaystyle}
\newcommand{\spur}[1]{\not\! #1 \,}
\begin{document}

\title{\hfill 
$\mbox{\small{\begin{tabular}{r}
${\rm Bari-TH/99-330}$\\
${\rm Napoli-DSF-99-2}$
%${\rm hep-ph/yymmxxx}$
\end{tabular}}}$ \\[1truecm]
Semileptonic and rare $B$ meson decays\\
into a light pseudoscalar meson}

\author{M. Ladisa$^{a,b}$, G. Nardulli$^{a,b}$, P. Santorelli$^{c,d}$} 

\date{{}}
\maketitle
\thispagestyle{empty}

\begin{it}
\begin{center}
$^a$Dipartimento di Fisica dell'Universit\`a di Bari, Italy\\
$^b$Istituto Nazionale di Fisica Nucleare, Sezione di Bari, Italy\\
$^c$Dipartimento di Scienze Fisiche, Universit\`a "Federico II" di Napoli, 
Italy\\
$^d$Istituto Nazionale di Fisica Nucleare, Sezione di Napoli, Italy
\end{center}
\end{it}

\begin{abstract}
In the framework of a QCD relativistic potential model we evaluate the 
form factors describing  the exclusive decays $B \to \pi \ell \nu$
and $B\to K \ell^+ \ell^-$. The present calculation extends a previous analysis 
of B meson decays into light vector mesons. We find results in agreement
with the data, when available, and with the theoretical constraints
imposed by the Callan-Treiman relation and the infinite heavy quark
mass limit.
\end{abstract}
\newpage

%\section{Introduction}
The study of the decays 
\be
B \to \pi \ell \bar \nu_{\ell}\\
\label{semilep}
\ee
\be
B \to K \ell^+ \ell^-
\label{rare}
\ee
represents a significant part of the experimental programmes at the next
proton-proton accelerators and at the future $B$-factories at SLAC and
KEK. 
The importance of these processes arises from the following reasons. The
decay (\ref{semilep}) allows to measure the product of the
Kobayashi-Maskawa ($KM$) matrix element $V_{ub}$ and the form factor
describing the decay process\footnote{There is one form factor
contributing to (\ref{semilep}) for a massless lepton.}
; similarly, the
decay (\ref{rare}) will give access, in appropriate regions of phase
space, to the $KM$ matrix element $V_{ts}$; therefore these processes
would allow to measure fundamental parameters of the {\it Standard 
Model} ($SM$) of the fundamental interactions, to say nothing of the
possibility to explore, in both cases, new effects beyond the $SM$. 

\par

It is fair to say, however, that, in spite of the fundamental relevance
of the processes (\ref{semilep}) and (\ref{rare}), the basic theory of
the hadronic  interactions, Quantum-Chromo-Dynamics (QCD), is still
unable to produce clear predictions for the hadronic matrix elements $B
\to \pi$, $B \to K$ involved in these decays. This is due to the lack of
a theoretical tool, as powerful as perturbation theory, able to produce
predictions for the nonperturbative quantities involved in these
processes. The most frequently used theoretical methods to deal with these 
problems are based on approximation  schemes
such as lattice QCD or QCD sum rules. These approaches have however
their own limitations. In the former method the finite lattice size
introduces a cut-off in the small momenta, which precludes the
possibility  to make reliable predictions in the small momentum transfer
region ($Q^2 \leq 15\ GeV^2$) (for recent reviews of lattice QCD predictions
for $B$ into light meson transitions see e.g. \cite{lat}).
In the case of QCD sum rules
or their variant, light cone sum rules, the theoretical uncertainties
are dominated by the peculiar theoretical tools employed by this method
(criteria for stability, hierarchic role of the different
nonperturbative contributions parametrized by the various condensates)
and cannot be reduced by adding new terms in the Operator Product
Expansion (for a discussion see \cite{Shifman}). 

\par

On the basis of these considerations, in \cite{Brho} 
we have presented an analysis of semileptonic and rare
transitions between the meson $B$ and a light vector meson in a QCD
relativistic potential model. In \cite{Brho} we argued that, because of
its simplicity, this model might be used as a viable alternative to the
more fundamental, but still limited theoretical approaches we have
discussed above. It is the aim of this paper to extend this analysis
to the decays (\ref{semilep}) and (\ref{rare}). 

\par
\noindent

To begin with, we review the main features of the QCD relativistic
potential model. It is a $potential\ model$ because the mesons are
described as bound states of constituent quarks and antiquarks tied
by an instantaneous potential $V(r)$. It is a $QCD\ model$ because the
potential is modelled according to the theory of the hadronic
interactions, i.e. it has a confining linear behaviour  at large
interquark distances $r$ and a Coulombic behaviour $\simeq -\alpha_s
(r)/r$ at small distances, with $\alpha_s (r)$ the running strong
coupling constant: in practice the interpolating Richardson's potential
$V(r)$ is used \cite{Rich}, cut-off at very small distances (of the
order of the inverse heavy meson mass) to take care of unphysical
singularities introduced by the relativistic kinematics \cite{Cea}.
Finally it is a $relativistic\ model$ because the wave equation used to
obtain the meson wave function $\Psi$ is the Salpeter equation embodying
the relativistic kinematics: 
\be
\left [ \sqrt{ -\nabla^2 + m^2_1} + \sqrt{ -\nabla^2 + m^2_2}
+V(r)\right ] \Psi(\vec r)=M \Psi(\vec r)~ \label{salp} \;\; ,
\ee
where, for heavy mesons made up by a heavy quark $Q$ and a light antiquark, 
1 refers the heavy quark and 2 to the light antiquark. The relativistic 
kinematics plays an important role when at least one of the two quarks 
constituting the meson is light, as in our case, and represents an improvement 
in comparison with the approach based on the non-relativistic quark model. 
In (\ref{salp}) $M$ is the heavy meson 
mass that is obtained by fitting the various parameters of the model, in 
particular the b-quark mass, that is fitted to the value $m_b=4890$ MeV, 
and the light quark masses 
$m_u\simeq m_d=38$ MeV, $m_s=115$ MeV
\footnote{Data on the heavy meson spectra are not of great help in fitting 
light quark masses, which, therefore, are not accurately determined in the 
model; its predictions, however, are not sensitive to $m_u$, $m_d$, $m_s$ 
values in most of the available kinematical range.}
. 
The $B$-meson wave function in 
its rest frame 
is obtained by solving (\ref{salp}); a useful representation in the momentum 
space was obtained in \cite{Brho} and is as follows
\be
\psi (k)=4\pi\sqrt{m_B \alpha^3}\ e^{-\alpha k} \;\; ,
\label{wf}
\ee
with $\alpha=2.4$ GeV$^{-1}$ and $k=|\vec k|$ the quark momentum in the 
$B$ rest frame.

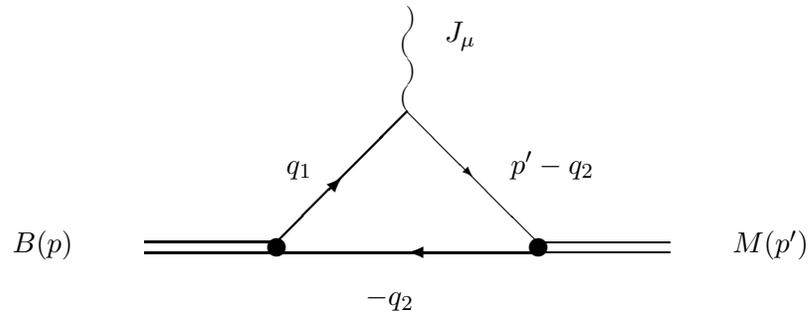
\begin{figure}[ht]
\begin{center}
\input FEYNMAN
\begin{picture}(15000,15000)
\THICKLINES
\drawline\fermion[\E\REG](0,0)[5000]
\drawline\fermion[\W\REG](14900,0)[9900]
\drawarrow[\LDIR\ATTIP](\pmidx,\pmidy)
\put(8000,-2000){ $-q_2$}
\put(5000,3000){ $q_1$}
\put(13500,3000){ $p^\prime-q_2$}
\drawline\fermion[\E\REG](0,400)[5000]
\put(-5000,0){$B(p)$}
\put(5000,200){\circle*{700}}
\drawline\fermion[\NE\REG](5000,400)[7000]
\drawarrow[\LDIR\ATTIP](\pmidx,\pmidy)
\THINLINES
\drawline\fermion[\SE\REG](\pbackx,\pbacky)[7000]
\drawarrow[\LDIR\ATTIP](\pmidx,\pmidy)
\drawline\photon[\N\REG](\pfrontx,\pfronty)[4]
\put(14900,200){\circle*{700}}
\put(11000,8000){ $J_\mu$}
\drawline\fermion[\E\REG](14900,0)[5000]
\drawline\fermion[\E\REG](14900,400)[5000]
\put(21900,0){ $M(p^\prime)$}
\end{picture}
\vskip 0.3 cm
\caption{Quark loop diagram describing the matrix element
$<M(p^{\prime})|J^{\mu}|B(p)>$; M is a light pseudoscalar meson,
$J_\mu=\bar q^{\prime} \Gamma^{\mu}b$ is the current inducing the decay and
$\Gamma^{\mu}$ is a combination of Dirac matrices.}
\label{f:fig3}
\end{center}
\end{figure}

The constituent quark picture used in the model is well suited for the
mesons comprising at least a heavy quark; for light mesons other
dynamical features, not accounted for by this simple picture, should be
incorporated, e.g. the nature of pseudo Nambu-Goldstone bosons of
$\pi$'s and $K$'s and the presence of important $spin-spin$ terms in
$V(r)$, not included in the Richardson's potential (their neglect for
heavy mesons is justified by the spin symmetry of the Heavy Quark
Effective Theory (HQET)\cite{HQET} which is valid  in the limit $m_Q \to
\infty$). The solution  adopted in \cite{Brho} was to avoid, for  light
mesons, the constituent quark picture and to describe their couplings to
the quark degrees of freedom by effective vertices. This assumption
produces a set of rules that are  used to compute the quark loop of fig.
\ref{f:fig3}, i.e. the diagram by which the hadronic amplitudes
describing the decays (\ref{semilep}) and (\ref{rare}) are evaluated.
They are as follows. 
\par\noindent
1) For a light pseudoscalar meson $M\ (=\pi^{\pm},K)$ of momentum 
$p^{\prime}$ we write the coupling 
\be
-\frac{N_q\ N_{q^{\prime}}}{f_M} \spur{p^{\prime}} \gamma_5 \;\;,
\label{0-}
\ee
where $f_M=f_{\pi}=130\ MeV$ or $f_M=f_K=160\ MeV$. The normalization factors 
$N_q,\ N_{q^{\prime}}$ for the quark coupled to the meson are discussed below. 

\noindent
2) For the heavy meson  $B$ in the initial state one introduces the
matrix:
\be
B=\frac{1}{\sqrt3}\psi (k){\sqrt{ \frac{m_q m_b}
{m_q m_b+q_1\cdot q_2} }}\;\;\frac{\spur{q_1}+m_b}{2 m_b}(-i\ \gamma_5)
\frac{-\!\!\!\spur{q_2}+m_q}{2 m_q}
\label{B}
\ee
where $m_b$ and $m_q$ are the heavy and light quark masses, $q^\mu_1,\
q^\mu_2$ their $4-$momenta. The normalization factor corresponds to the
normalization $<B|B>=2\ m_B$ and $\dd\int \frac{d^3 k}{(2\pi)^3}
|\psi(k)|^2=2 m_B$ already embodied in (\ref{B}). One assumes that the
$4-$momentum is conserved at the vertex $B\bar qb$, i.e.
$q^\mu_1+q^\mu_2=p^\mu=$ B meson $4-$momentum. Therefore
$q^\mu_1=(E_b,\vec k),~q^\mu_2=(E_q,-\vec k)$ and 
\be
E_b+E_q=m_B \;\; .
\label{Alt-Cab}
\ee
3) To take into account the off-shell 
effects due to the quarks interacting in the meson, one introduces running 
quark masses $m(k)$, to enforce the 
condition
\be
E=\sqrt{m^2(k)+|\vec k|^2}
\label{Alt-Cab2}
\ee
for the constituent quarks. For the kinematics of the decays (\ref{semilep}) 
and (\ref{rare}) it is sufficient to introduce the running mass only for the 
heavy quark
\footnote{By this choice, the average $<m_b(k)>$ does not differ 
significantly from the value $m_b$ fitted from the spectrum, see \cite{Brho} 
for details.}
\be
m_b=m_b(k)\;\; ,
\label{running}
\ee
defined by the condition
\be
\sqrt{m_q^2+|\vec k|^2} + \sqrt{m_b^2+|\vec k|^2} = m_B \;\; .
\label{Alt-Cab3}   
\ee
\noindent
4) The condition $m_b^2 \geq 0$ implies the constraint 
\be
0\leq k\leq k_M=\frac{m_B^2-m^2_q}{2 m_B}~,\label{kmax}
\ee
on the integration over the loop momentum $k$
\be
\int\frac{d^3k}{(2\pi)^3}\;\; .
\label{loop}
\ee
5) For each quark line with momentum $q$ and not representing a constituent 
quark one introduces the factor
\be
\frac{i}{\spur q-m_{q^{\prime}}}\times G(q^2)~,
\label{ff}
\ee
where $G(q^2)$ is a shape function that modifies the free propagation of the 
quark of mass $m_{q^{\prime}}$ in the hadronic matter. The shape function 
\be
G(q^2)=\frac{m^2_G-m^2_{q^{\prime}}} {m^2_G-q^2}
\label{ff2}
\ee
was  adopted in \cite{Brho}; the value of the mass parameter
$m_G$ was determined in \cite{Brho} by the experimental data on the $B \to K^*
\gamma$ decay.  A range $[1.2,7.6]$ GeV$^2$ of possible values
of $m_G^2$ was obtained.

\noindent
6) For the hadronic current in fig. \ref{f:fig3} one puts the factor 
\be
N_{q} N_{q^\prime}\Gamma^\mu\;,
\label{J}
\ee
where $\Gamma^\mu$ is a $4 \times 4$ matrix. We shall consider 
$\Gamma^\mu=\gamma^\mu$ and $\Gamma^\mu=\sigma^{\mu\nu}q_{\nu}$ 
(with $\sigma^{\mu\nu}=i/2\ [\gamma^{\mu},\gamma^{\nu}]$). The normalization 
factor $N_q$ is as follows:
\be
N_q= 
\left\{\begin{array}{ccl}
\dd\sqrt{\frac{m_q}{E_q}} &  ~~ &  {\rm(if ~q=constituent~ quark)} \\
 & \\
1 & ~~ & {\rm (otherwise)\, .}
\end{array}\right . 
\label{Nq}
\ee

\noindent
7) For each quark loop one puts a colour factor of 3 and performs a trace over 
Dirac matrices.

%\section{$B$ to Light Pseudoscalar Meson Transition Form Factors}

This set of rules  
can now be applied to the evaluation of the matrix element
$<M(p^{\prime})|\bar q^{\prime} \Gamma^{\mu} b|B(p)>$ with the result:
\bea
&& <M(p^{\prime})|\bar q \Gamma^{\mu} b|B(p)> =
\sqrt{3}\int\frac{d^3k}{(2\pi)^3}\theta[k_M-k]\psi(k)
\sqrt{\frac{m_q m_b}{m_q m_b+q_1\cdot q_2}} \nn\\
&&Tr\left [\frac{\spur{q_1}+m_b}{2 m_b}(-i\ \gamma_5)
\frac{-\spur{q_2}+ m_q}{2 m_q} \sqrt{\frac{m_b m_q}{E_b E_q}}
\frac{-1}{f_M}(\spur{p}-\spur{q}) \gamma_5
\ \frac{i\ G[(q_1-q)^2]}{\spur{q_1}-\spur{q}-m_{q^{\prime}} + i\ \epsilon}
\Gamma^{\mu} \right ]
\;\; .
\label{ampiezza}
\eea

\noindent
From this expression one can obtain the relevant formulae 
for the various form factors. 
With $q=p-p^{\prime}$, we write
\bea
<M(p^{\prime})|\bar q^{\prime} \gamma^{\mu} b|\bar B (p)> 
&=& f_+(q^2)(p+p^{\prime})^{\mu}+f_-(q^2)q^{\mu} \nn \\
&=& F_1(q^2)(p+p^{\prime})^{\mu} + \frac{m_B^2-m_M^2}{q^2}
q^{\mu}\left (F_0(q^2) - F_1(q^2) \right) \;\; ,
\label{B-M-sl}
\eea

%\bea
%<M(p^{\prime})|\bar q^{\prime} \gamma^{\mu} b|\bar B (p)> 
%&=& f_+(q^2)(p+p^{\prime})^{\mu}+f_-(q^2)q^{\mu} \nn \\
%&=& F_1(q^2) \left[(p+p^{\prime})^{\mu} - \frac{m_B^2-m_M^2}{q^2}
%q^{\mu}\right] \nn \\ 
%&+& F_0(q^2)\frac{m_B^2-m_M^2}{q^2} q^{\mu} \;\; ,
%\label{B-M-sl}
%\eea
%
\bea
<M(p^{\prime})|\ i\ \bar q^{\prime} \sigma^{\mu\nu}q_{\nu} b|\bar B(p)> &=&
\frac{f_T(q^2)}{m_B + m_M}
\left[(p+p^{\prime})^{\mu} q^2 - (m_B^2-m_M^2) q^{\mu}\right] \;\; ,
\label{B-M-rare}
\eea
where
\bea
F_1(q^2)   &=& f_+(q^2) \nn \\
F_0(q^2)   &=& f_+(q^2) + \frac{q^2}{m_B^2-m_M^2} f_-(q^2) \;\; .
\eea
In (\ref{B-M-sl}) and (\ref{B-M-rare}) we shall consider  $M=\pi$ or 
$M=K$ since both cases are of physical interest if we wish to
consider not only 
semileptonic and radiative transitions, but also nonleptonic decays. 

\noindent

The calculation of the trace and the integral in 
(\ref{ampiezza}) is straightforward and is similar to the one obtained in 
\cite{Brho} for $B \to \rho,\ B \to K^*$ transitions. For all the form factors 
we write 
$F(q^2)=F(q^2,m_{q^{\prime}})-F(q^2,m_G)$, where, for the various form 
factors, we have
\bea
F_0(q^2,x) &=& 
%\frac{q_{\mu}\ <M(p^{\prime})|\bar q^{\prime} \gamma^{\mu} b|\bar B(p)>}
%{m_B^2-m_M^2} = 
\frac{\sqrt{6}}{4 \pi^2 f_M (m_B^2-m_M^2)} 
\int_0 ^{k_M} \frac{dk\ \ k^2 \psi(k)}{\sqrt{E_q E_b [m_{B}^2-(m_b - m_q)^2]}}
\times \nn \\
&&\int_{-1} ^1 dz\ 
\frac{1}{m_M^2 - 2 E_q (m_B - q^0) + m_q^2 - x^2 + 2 \qv\ k\ z} \nn \\
&& \Bigg\{\ [2 E_q (m_B - q^0)-m_M^2-2 \qv\ k\ z]
   [(m_q\ E_b + m_b\ E_q)q^0 + (m_b - m_q)\ \qv\ k\ z] + \nn\\
&& \frac{m_q + m_{q^{\prime}}}{2}
\left[\ (q^2-m_B\ q^0)\ \left[ m_B^2 - (m_b - m_q)^2 \right] 
+ 2\ m_B^2\ \qv\ k\ z\ \right]\ \Bigg\} \;\; ,
\label{fdfF0}
\eea

\bea
F_1(q^2,x) &=& \frac{\sqrt{6}}{8 \pi^2 f_M}
\int_0 ^{k_M} \frac{dk\ \ k^2 \psi(k)}{\sqrt{E_q E_b [m_{B}^2-(m_b - m_q)^2]}}
\times \nn \\
&&\int_{-1} ^1 dz\
\frac{1}{m_M^2 - 2 E_q (m_B - q^0) + m_q^2 - x^2 + 2 \qv\ k\ z} \nn \\
&&\Bigg\{\ [2 E_q (m_B - q^0)-m_M^2-2 \qv\ k\ z]
\frac{\qv (m_q E_b + m_b E_q) + q^0\ k\ z\ (m_b - m_q)}{m_B\ \qv} + \nn \\
&& \frac{m_q + m_{q^{\prime}}}{2}\  
\frac{2\ k\ z\ (m_B\ q^0 - q^2) - \qv\ \left[m_B^2 - (m_b - m_q)^2\right]}
{\qv}\ \Bigg\} \;\; ,
\label{fdfF1}
\eea

\bea
f_T(q^2,x) &=& -\frac{\sqrt{6}}{4 \pi^2 f_M}\ \frac{m_B + m_M}{2}
\int_0 ^{k_M} \frac{dk\ \ k^2 \psi(k)}{\sqrt{E_q E_b [m_{B}^2-(m_b - m_q)^2]}}
\times \nn \\
&&\int_{-1} ^1 dz\
\frac{1}{m_M^2 - 2 E_q (m_B - q^0) + m_q^2 - x^2 + 2 \qv\ k\ z} \nn \\
&&\Bigg\{\ [2 E_q (m_B - q^0)-m_M^2 - 2 \qv\ k\ z]\ \frac{k\ z}{\qv} + \nn \\
&& (m_q + m_{q^{\prime}}) 
\frac{ -(m_b E_q + m_q E_b)\ \qv + (m_B -q^0)\ k\ z\ ( m_b - m_q)}{m_B\ \qv}\ 
\Bigg\} \; .
\label{fdffT}
\eea
In these equations $q_0$ is the time component of four-momentum $q^\mu$,
\be
z = cos(\theta) \;\; ,
\ee
with $\theta$ the angle between $\vec k$ and the direction of
transferred momentum $\vec q$. We note that, for $M=\pi$,
$m_q=m_{q^{\prime}}=m_u$ and $f_M=f_{\pi}$, while, for $M=K$,
$m_q=m_u$, $m_{q^{\prime}}=m_s$ and $f_M=f_K$. 

\par
\noindent 

Before discussing our numerical results in detail let us compute
$F_0(q^2)$ for $q^2=m_B^2-m_M^2$; in the chiral limit $F_0(m_B^2-m_M^2)
\simeq F_0(m_B^2)$ must obey the $Callan-Treiman$ relation
\cite{CallanTreiman} 
\be
F_0(m_B^2)=\frac{f_B}{f_{\pi}} \;\; .
\label{Call-Tre}
\ee
This is therefore a consistency test to be satisfied by the model.
We have numerically evaluated
$F^{B\pi}_0(m_B^2)$ for different values of
the parameter $m_G$ and we have obtained the result
$F^{B\pi}_0(m_B^2) \simeq 1.48$, almost independent of
$m_G$. This result should be compared to $f_B/f_{\pi} \simeq
1.58$, which is obtained using  $f_B=0.2\ GeV$, 
i.e. the value computed  in \cite{Brho} using the present 
model. The small discrepancy in the Callan-Treiman relation 
($\simeq 6$\%) may be attributed to the deviations induced in the
$B$ meson wave function by the chiral limit that are not accounted
for by this calculation. We expect however that these differences 
vanish if, in addition to the chiral limit, one also takes the
infinite heavy quark mass limit; as a matter of fact one can verify 
rather easily, using the previous formula for $F_0$ and the expression
in \cite{Brho} for $f_B$, that the Callan-Treiman exactly holds  in the
combined $m_b \to \infty$ and $m_M\to 0$ limit.

Let us now consider the form factor $F_1(q^2)$ (respectively $f_T(q^2)$). 
Our numerical results for the central value of $m_G$, i.e. $m_G=1.77$ GeV, 
show that the $q^2-$behaviour of
this form factor is increasing (resp. decreasing)
 for both small and moderate values
of $q^2$, independently of the value of the mass parameter $m_G$
introduced in eq.(\ref{ff}). This behaviour should hold also at
large $q^2$ 
($q^2 \geq 15\ GeV^2$) due to the effect, in this region, of a pole in  
the $q^2$ functional dependence, predicted by the
 dispersion relation. Differently 
from our analysis of $F_0(q^2)$,  we 
cannot pretend, however, to extend the validity of our predictions for
 $F_1(q^2)$ and $f_T(q^2)$ at the extreme values of $q^2$. The difference
between the two cases is as follows. In the case of $F_0$, 
the pole (with $J^P=0^+$) contribution to this form factor vanishes 
in the chiral limit and has therefore a minor impact on the  $q^2$ 
behaviour. On the contrary the form factors $F_1$ and $f_T$, 
have a non vanishing polar contribution which
becomes larger and larger with increasing $q^2$. While we expect that
this behaviour become visible well before the pole, 
at extreme values of $q^2$ the diverging behaviour induced by such a 
contribution cannot be reproduced by the model.  As a matter of fact,
for larger values of $q^2$, $\qv$ becomes 
smaller and smaller, and, therefore, the model 
becomes sensitive to the actual values of the parameters, in particular the
light quark masses that are not accurately fitted by the
available experimental data (see above). 
Therefore we can consider that our predictions are reliable in the range 
$(0,\ 15)$ GeV$^2$; at $q^2=0$ we get 
\bea
F^{B \pi}_0(0)&=&F^{B \pi}_1(0) = 0.37 \pm 0.12 \nn\\
F^{B K}_0(0)&=&F^{B K}_1(0) = 0.26 \pm 0.08 \nn\\
f_T^{B \pi}(0)&=&-0.14 \pm 0.02 \nn\\ 
f_T^{B K}(0)&=&-0.09^{+0.05}_{-0.02} \; .
\label{results}
\eea
The central values are obtained for $m_G=1.77$ GeV, which is
the best fit of the parameter $m_G$ found
in \cite{Brho} by the experimental branching ratio 
${\cal B}(B \to K^* \gamma)$, whereas the theoretical uncertainty is obtained 
by varying $m_G$ in the range $[1.1,\ 2.8]$ GeV. The results for $B \to \pi$ 
refer to charged pions.

\noindent

\begin{figure}[t]
\begin{center}
\epsfig{file=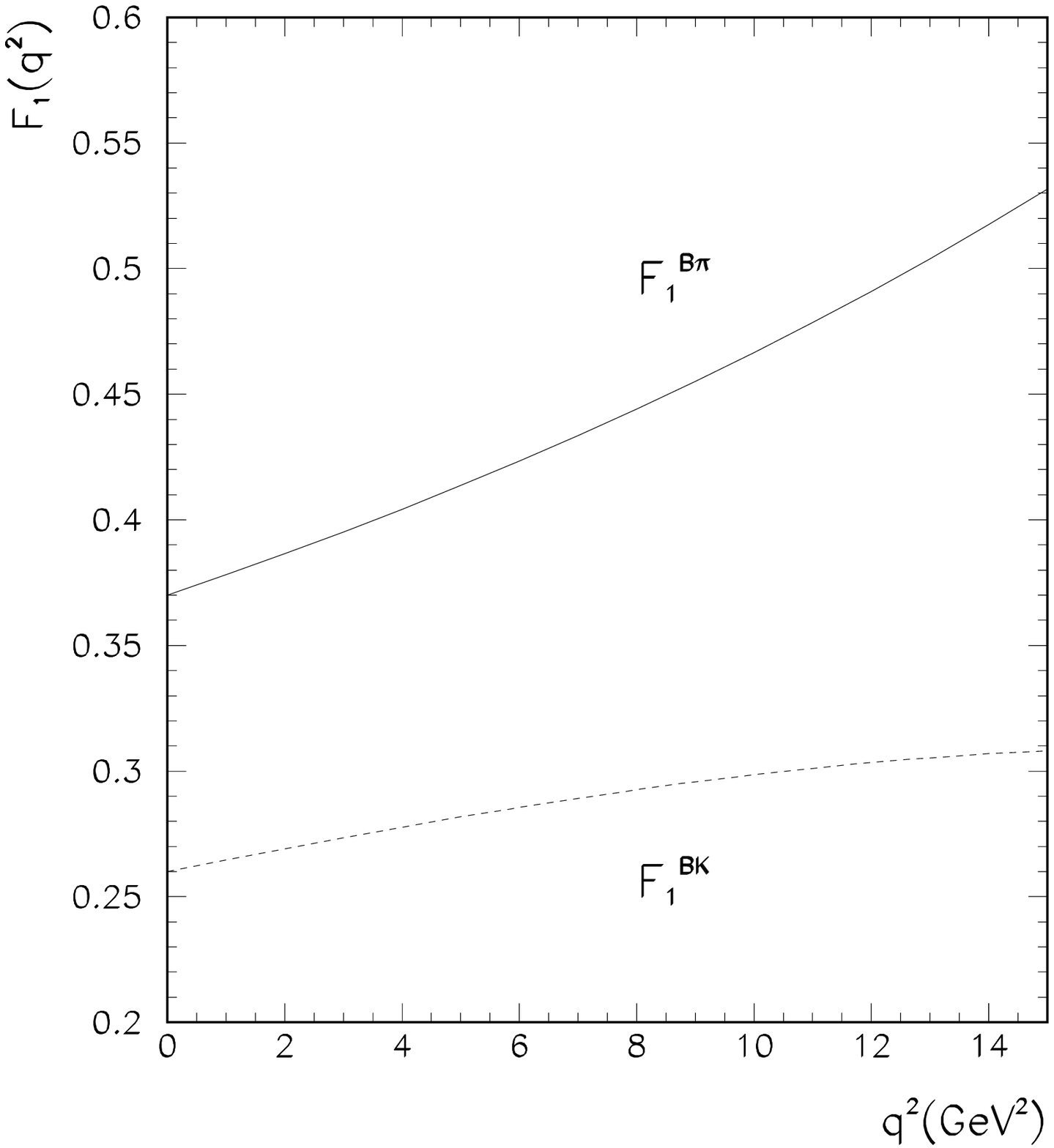,height=8cm} \quad
\epsfig{file=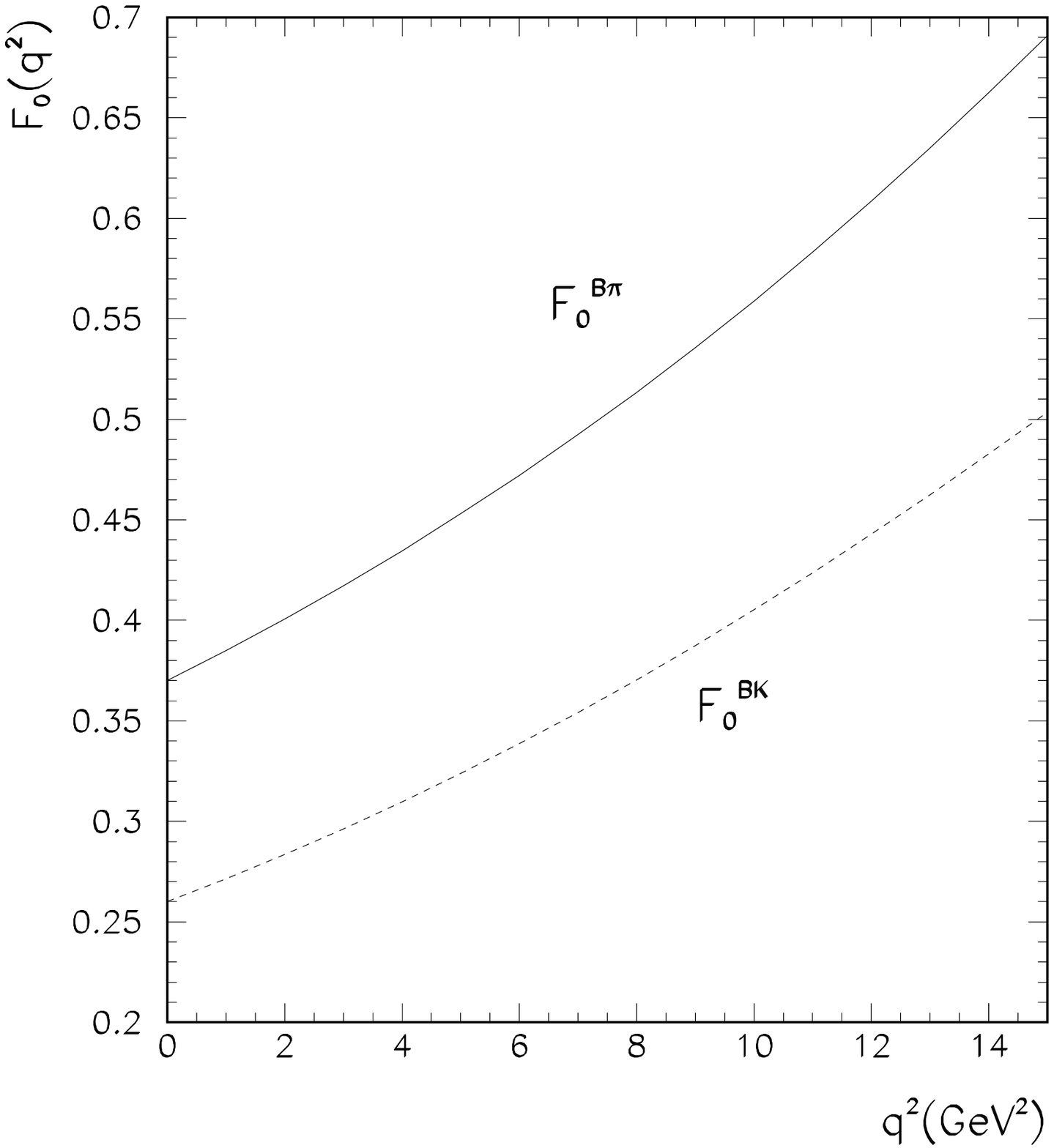,height=8cm}
\end{center}
\caption{$F_0(q^2),\ F_1(q^2)$ for $B \to \pi$ and $B \to K$ 
transitions.} 
\label{f:fig1}
\end{figure}

\begin{figure}[t]
\begin{center}
\epsfig{file=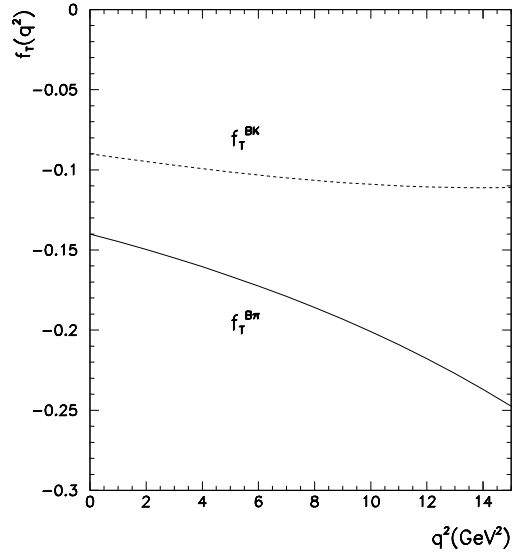,height=8cm}
\end{center}
\caption{$f_T(q^2)$ for $B \to \pi$ and $B \to K$ transitions.}
\label{f:fig2}
\end{figure}

Let us now consider the $q^2$-behaviour of the form factors. We introduce 
the two-parameter function for the three form factors 
\be
F(q^2)=\frac{F(0)}{1~-~a_F \left(\dd\frac{q^2}{m_B^2}\right) +~b_F 
\left(\dd\frac{q^2}{m_B^2}\right)^2}\;\; ;
\label{F-param}
\ee
\noindent
here $a_F~,b_F~$ are parameters to be fitted by  means of the
numerical analysis and $F(0)$ is given in eq. (\ref{results}); 
to allow a comparison with other approaches 
we perform the analysis up to $q^2=15$ GeV$^2$, 
both for $\pi$
and $K$ mesons. We collect the fitted values in table \ref{t:tab1} and report 
the $q^2$-dependence in fig. \ref{f:fig1}, \ref{f:fig2}.

\begin{table}[ht!]
\begin{center}
\begin{tabular}{||c||c|c|c||c|c|c|c||}
\hline\hline
& $F(0)$ & $a_F$ & $b_F$ & $F(0)$ & $a_F$ & $b_F$ & \\
\hline \hline
$F_1^{B \pi}$  &
$0.37$ & $0.60$ & $0.065$ &
$0.26$ & $0.50$ & $0.39$ &
$F_1^{B K}$\\
$F_0^{B \pi}$ &
$0.37$ & $1.1$ & $0.44$ &
$0.26$ & $1.2$ & $0.56$ &
$F_0^{B K}$\\
$f_T^{B \pi}$ &
$-0.14$ & $0.92$ & $0.21$ &
$-0.09$ & $0.76$ & $0.76$ &
$f_T^{B K}$\\
\hline \hline
\end{tabular}
\caption{
Parameters appearing in eq. (\ref{F-param}) for different 
$B$ form factors.}
\label{t:tab1}
\end{center}
\end{table}

\par
\noindent
From table \ref{t:tab1} and from fig. \ref{f:fig1}, \ref{f:fig2}, one can see 
that $F_0(q^2)$, $F^{B \pi}_1(q^2)$ and $f^{B \pi}_T(q^2)$ 
have a $q^2$ behaviour similar to a
 single pole. For $F^{B K}_1(q^2)$ and $f^{B K}_T(q^2)$ there are significant 
deviations from this behaviour. We do not have yet experimental 
data to test these predictions and we shall limit to compare our results with 
other theoretical approaches; before doing that, let us discuss the infinite 
heavy quark mass limit of the model.
In the framework of the Heavy Quark Effective Theory, which corresponds to the 
$m_b \to \infty$ limit, there is a constraint to be satisfied by the three 
form factors, i.e., the relation originally found in \cite{Isgur-Wise}:
\be
f_T(q^2)=-\frac{m_B+m_M}{2\ m_B} \left[ F_1(q^2)- \left( m_B^2-m_M^2 \right)
\frac{F_0(q^2)-F_1(q^2)}{q^2} \right] \;\; .
\label{Isgur-Wise}
\ee
This relation holds  in the limit $m_b \to \infty$ and for high
$q^2$ ($q^2\simeq q^2_{max}$). We have checked that this relation formally 
holds in our model in the infinite heavy quark mass limit and for 
$q^2\simeq q^2_{max}$. For 
the actual value of $m_B$ ($=5.28$ GeV) and for the transition $B \to \pi$ the 
situation is as follows. We choose $q^2_M=15$ GeV$^2$, i.e. the maximum
value at which we can trust our predictions and we find numerically
$F_1(q^2_M)\simeq 0.54,~ F_0(q^2_M)\simeq 0.71,~f_T(q^2_M)\simeq
-0.27$. Therefore 
the relation (\ref{Isgur-Wise}) has a significant 
violation of $50\%$, that may be attributed to the fact that we are still
far from $q^2_{max}$ and ${\cal O}(1/m_b)$ corrections are large. 
Similar results
are obtained for the $B\to K$ transition.

Let us finally comment on the scaling laws of the form factors at large $q^2$
that can be helpful in using the heavy flavour symmetry to relate the form
factors of $B$ and $D$ mesons \cite{casalbuoni}.   
From eqs.(\ref{fdfF1}) and (\ref{fdffT}) the following
behaviours can be formally derived:
\begin{eqnarray}
F_1(q^2_{max}) & \approx & \sqrt{m_b} \label{ff1}\\
f_T(q^2_{max}) & \approx & \sqrt{m_b}~;\label{fft}
\end{eqnarray}
they are in agreement with the pole (vector meson) dominance of the 
form factors observed in \cite{wise}
\cite{casalbuoni}; moreover in the chiral limit,
from eq.(\ref{fdfF0}) one gets:
\begin{equation}
F_0(q^2_{max})\approx \frac{1}{\sqrt{m_b}}\label{ff0} \;\; .
\end{equation} 

\par

Let us now compare our work with other theoretical approaches.
In table \ref{t:tab2} we compare our outcome for the values at 
$q^2=0$ with the results of QCD sum rules and lattice QCD calculations 
(for other work on this subject see, e.g. \cite{Belyaev}).
We observe that our results are in agreement, within the theoretical 
uncertainties, with the 
determinations obtained by light cone sum rules 
$(LCSR)$ \cite{Ball}, lattice \cite{lat} 
and lattice $+\ LCSR$ \cite{DelDebbio}.  

\begin{table}[ht!]
\begin{center}
\begin{tabular}{||c||c|c|c||c|c||}
\hline \hline
& $This~work$ &
$LCSR$ \cite{Ball} &
$SR$ \cite{noi} &
$Latt.$ \cite{lat} &
$\begin{array}{c}
Latt.~+ \\
$LCSR$\ \cite{DelDebbio}
\end{array}$ \\
\hline \hline
$F_1^{B \pi}(0)$   & 
$0.37 \pm 0.12$  & $0.30 \pm 0.04$  & $0.24$  & $0.27 \pm 0.11$  & 
$0.27 \pm 0.11$ \\
$f_T^{B \pi}(0)$ &
$-0.14 \pm 0.02$  & $-0.19 \pm 0.02$ & $--$  & $--$ & $--$  \\
\hline \hline
$F_1^{B K}(0)$   & 
$0.26 \pm 0.08$  &$0.35 \pm 0.05$ & $0.25$  & $--$ & $--$  \\ 
$f_T^{B K}(0)$ &
$-0.09^{+0.05}_{-0.02}$  & $-0.15 \pm 0.02$ & $-0.14$  & $--$  & $--$ \\
\hline \hline
\end{tabular}
\caption{Comparison of the results coming from different approach 
to evaluate form factors.}
\label{t:tab2}
\end{center}
\end{table}

\par
\noindent

As for the $q^2$ dependence, we have not reported the predictions 
of other theoretical approaches, since they qualitatively agree with our 
calculations. 
In absence of detailed experimental data on the form factors,
the best we can do to test the model is to
use data on the partial width $\Gamma(B \to \pi \ell \nu)$.
To perform this comparison we must, however, extrapolate the $q^2$-behaviour 
obtained by eq. (\ref{F-param}) and tables \ref{t:tab1} and \ref{t:tab2}, and 
valid in the region $(0,15)$ GeV$^2$, to the whole $q^2$ range. This procedure 
implies an uncertainty which is difficult to assess, but should not be 
extremely large due to the phase space limitation at high $q^2$.
We obtain 
\be
{\cal BR}(\bar B^0 \to \pi^+ \ell^- \bar \nu) =
1.03\ \left(\frac{|V_{ub}|}{3.2~10^{-3}}\right)^2\ 10^{-4}~,
\ee to be compared to 
the experimental value
${\cal BR}(B \to \pi \ell \nu)_{exp.}= ( 1.8 \pm 0.6 )\ 10^{-4}$ \cite{PDG}. 
Therefore our result is compatible with
the present range of the $KM$ matrix element $V_{ub}=[1.8,\ 4.5]\times 10^{-3}$
; the preferred range of values selected by the 
model and by the present experimental limits on $V_{ub}$ is 
$V_{ub}=(4.0 \pm 0.5)\times 10^{-3}$. 

We conclude our analysis by summarizing our results. We have used a 
QCD relativistic potential model, introduced in \cite{Brho}, 
to study the weak 
and radiative transitions $B\to \pi,~K$. 
We have computed the relevant form factors
and tested the Callan-Treiman and the Isgur-Wise relation. The former
relation, valid in the chiral limit, is satisfied at the 6$\%$ level,
while the latter, valid in the $m_b \to \infty$ limit, 
has significant violations,  due to ${\cal O}(1/m_b)$ corrections. Our
result for the branching ratio 
${\cal BR}(B \to \pi \ell \nu)$ agrees with the experimental data.  

\par
{\bf Aknowledgement}
We thank P. Colangelo and F. De Fazio for the collaboration in the early stage 
of this work.

\newpage

\end{document}